\documentclass[letterpaper]{article} 
\usepackage{aaai2027}  
\usepackage[hyphens]{url}  
\usepackage{graphicx} 
\usepackage{natbib}  
\usepackage{caption} 
\usepackage{algorithm}
\usepackage{makecell}
\usepackage{algorithmic}
\usepackage{graphicx}
\usepackage{amsmath,amssymb,amsfonts}
\usepackage{algorithmic}
\usepackage{graphicx}
\usepackage{textcomp}
\usepackage{xcolor}
\usepackage{tabularx}
\usepackage{booktabs}
\usepackage{array}
\usepackage{threeparttable}
\def\BibTeX{{\rm B\kern-.05em{\sc i\kern-.025em b}\kern-.08em
    T\kern-.1667em\lower.7ex\hbox{E}\kern-.125emX}}
\usepackage{newfloat}
\usepackage{listings}
\DeclareCaptionStyle{ruled}{labelfont=normalfont,labelsep=colon,strut=off} 
\floatstyle{ruled}
\newfloat{listing}{tb}{lst}{}
\floatname{listing}{Listing}

\usepackage{booktabs}

\title{AgentForge: An Immersive Role-Playing Platform for Learning Agentic Software Engineering}

\author{
    Zihan Fang,
    Yueke Zhang,
    Yu Huang
}

\affiliations{
    Vanderbilt University\\
    Nashville, Tennessee, USA\\
    \{zihan.fang, yueke.zhang, yu.huang\}@vanderbilt.edu
}

\begin{document}

\maketitle

\begin{abstract}
Agentic AI is increasingly used to coordinate planning, implementation, review, and testing in software development, yet it often offers limited transparency into its decisions and interactions. Many such systems also assume that users can effectively guide the AI's decisions and validate its outputs. This assumption poses a particular challenge for novices, who must simultaneously learn how agentic AI works, how to collaborate with it effectively, and how to evaluate its outputs critically.
To address this challenge, we present \textit{AgentForge}, an immersive learning system in which novices take on one of four software-engineering roles: Task Planner, Patch Author, Code Reviewer, or Test Runner, within a multi-agent code-repair workflow. In each practice session, the novices perform their chosen role while AI agents perform the remaining three. 
Through role-based scaffolding and metacognitive support, AgentForge clarifies role-specific responsibilities, makes agent coordination and intermediate artifacts visible, and encourages novices to monitor and evaluate their decisions.
In a study with 37 novice developers, participants achieved high task-completion rates with AI-agent support. 
However, interaction demands differed significantly across practices: the Code Reviewer practice required more interaction turns, reroutes, and completion time ($p_{\mathrm{adj}} = .004$) and was perceived as the most challenging.
Participants nevertheless reported significant gains in their understanding of software repair and agent collaboration ($p_{\mathrm{adj}} < .001$). These findings suggest that AgentForge can help novices develop practical software-engineering skills while learning to collaborate with agentic AI more critically and effectively.
\end{abstract}

\section{Introduction}
Advances in large language models (LLMs) have rapidly transformed how software is learned and developed, both in the classroom and real-world settings~\cite{shahzad2025comprehensive,annepaka2025large}. 
More recently, agentic AI in software engineering has extended LLM assistance beyond single-turn code completion by organizing development activities into goal-directed workflows, in which multiple agents coordinate planning, implementation, testing, and iterative refinement~\cite{bandi2025rise}. 
Under this paradigm, the programmer's role shifts from writing code directly to interpreting intermediate artifacts, validating agent outputs, and guiding the decisions across the workflow.
Building on this shift, recent multi-agent frameworks have improved the quality of generated code through role-based collaboration. 
For example, ChatDev organizes development into specialized phases to improve completeness, executability, and consistency with requirements~\cite{qian2024chatdev}. MetaGPT coordinates agents through standardized operating procedures~\cite{hong2024metagpt}, whereas AgentCoder assigns programmer, test-designer, and test-executor roles to iteratively generate, test, and refine code~\cite{huang2023agentcoder}.

However, despite their ability to support complex software development workflows, many agentic AI systems assume that users can already decompose tasks, guide agents, and validate outcomes, while providing limited visibility into how those agents make decisions and coordinate. 
Accordingly, prior work identifies competencies in software engineering, AI interaction, adjacent technical domains, and nontechnical skills such as problem framing and critical thinking as important for effective use~\cite{kam2025professional}.
This is especially important for novices, who must understand agent roles and coordination and learn to evaluate whether generated outputs are correct and maintainable~\cite{he2025llm,wang2025devcoach,kazemitabaar2025exploring,pezze20252030}. Without appropriate instructional support, they may passively accept AI outputs rather than reason about their quality, limiting their ability to develop the judgment needed for effective human–AI collaboration in software development.

To address these challenges, we designed \textit{AgentForge}, an immersive learning system in which novices actively assume software-engineering roles within a multi-agent workflow.
With scaffolding\footnote{Scaffolding refers to structured support that guides novices through tasks they may not yet be able to complete independently.} and metacognitive support\footnote{Metacognitive support refers to guidance that helps novices monitor, evaluate, and regulate their understanding and decision-making.}, novices observe agent coordination, interpret intermediate artifacts, guide AI decisions, and evaluate AI-generated outcomes.
We focus on code repair, an authentic software-engineering activity involving problem understanding, code modification, review, and validation. 
\textit{AgentForge} decomposes this process into four roles: Task Planner, Patch Author, Code Reviewer, and Test Runner. 
In each practice, novices assume one role while AI agents perform the others. We recruited 37 novice developers to examine how they interact with agentic AI during code-repair practice, their task performance, and changes in their understanding of software repair and agent collaboration.
We found that participants can achieve high task-completion rates with AI-agent support. 
However, interaction demands differed significantly across practices: the Code Reviewer practice required more human–AI interactions and reroutes and had longer completion times ($p_{\mathrm{adj}} = .004$), and was most often perceived as the most challenging, given its nuanced review and validation decisions.
Nevertheless, participants reported significant gains in their understanding of both software repair and agentic AI collaboration (both $p_{\mathrm{adj}} < .001$).
We claim the following contributions:
\begin{itemize}
\item An immersive learning environment that simulates role-based software-engineering workflows for novice developer training.
\item A scaffolding design that builds novices' practical software-engineering skills and critical AI collaboration.
\item Empirical findings on novices' metacognitive engagement with agentic AI during software development.
\end{itemize}

\section{Related Work}

\subsection{AI-Assisted Programming for Novice Developers}
Advances in LLMs, such as ChatGPT, GitHub Copilot, and Codex, have created new opportunities and challenges for novice developers learning programming and real-world software development~\cite{ferino2025novice, tabarsi2025llms, fu2025large}. Prior work shows that novices use generative AI to understand code, complete assignments, debug, and seek explanations~\cite{prather2023robots,scholl2024novice}. AI tools are therefore becoming an important part of computing education and practice, supporting personalized tutoring, code explanation, automated feedback, and assessment~\cite{raihan2025large, zhu2025systematic, fang2025dpo}.

At the same time, AI-assisted programming raises concerns about overreliance, shallow understanding, and fewer opportunities to practice core problem-solving skills~\cite{liu2026tool, alanazi2025influence}. Although AI can support engagement and help students produce code and test suites, careful design is needed to keep novices actively involved in reasoning about generated solutions~\cite{kazemitabaar2023studying,fang2025comparative}. Similarly, while novices often view AI tools as helpful, their benefits depend on how they are integrated into development and whether users critically evaluate AI-generated outputs~\cite{scholl2024novice,lepp2025does}.
However, most prior work examines AI’s effects on novices through isolated programming tasks rather than through structured, realistic software development practices. These limitations motivate systems that scaffold novices’ participation in authentic workflows instead of merely providing answers. Accordingly, our system uses a structured multi-agent workflow to teach not only AI-assisted coding, but also how development tasks are decomposed, coordinated, reviewed, and validated.

\subsection{Human-in-the-Loop AI in Software Engineering}
Many AI systems have been developed to support human-in-the-loop software engineering workflows~\cite{takerngsaksiri2025human,nascimento2018toward}, in which AI assists with development tasks while humans interpret its outputs, make decisions, and validate proposed changes~\cite{alenezi2025ai}.
For example, SWE-agent introduces an agent-computer interface that allows language-model agents to navigate repositories, edit files, and run tests to resolve real GitHub issues~\cite{yang2024swe}. AutoCodeRover similarly combines LLMs with code search and program analysis to address GitHub issues through autonomous program improvement~\cite{zhang2024autocoderover}. These systems reflect a shift from passive code suggestion toward active AI collaboration, though they still operate as a single agent handling all steps of the task, with human oversight remaining important for correctness, maintainability, and alignment with project goals.

Building on this shift toward active collaboration, growing research explores multi-agent software development, where specialized LLM-based agents coordinate across distinct roles to complete development tasks. ChatDev models software development as a virtual organization of role-specialized agents involved in design, coding, and testing~\cite{qian2024chatdev}. MetaGPT similarly encodes standard operating procedures into multi-agent collaboration, assigning roles such as product manager, architect, engineer, and project manager~\cite{hong2024metagpt}. These frameworks suggest that complex software-engineering tasks benefit from role decomposition, structured communication, and explicit workflow coordination. 
From a human-in-the-loop perspective, such structures also allow humans to intervene at specific workflow stages, inspect intermediate artifacts, and guide or revise AI-generated decisions~\cite{chappidi2026does}.
However, most multi-agent systems prioritize automation or benchmark performance, with limited attention to human participation and learning. Because agent reasoning, roles, and handoffs are often hidden, novice developers in particular struggle to understand how tasks are decomposed, decisions are made, and outputs are evaluated. Our human-in-the-loop system addresses this gap by having learners assume one agent role while AI agents perform the others, requiring novices to make role-specific decisions, interpret AI-generated outputs, and understand how those outputs affect the broader workflow.


\subsection{Immersive Learning for Software Development}
Immersive learning emphasizes active participation, situated experience, and engagement with authentic tasks~\cite{dawley2013situated,beck2023educational}. 
It aligns with experiential learning theory, which describes learning as a cycle of concrete experience, reflection, conceptualization, and active experimentation~\cite{kolb2014experiential}.

In software engineering education, immersive approaches help novices move beyond abstract concepts through realistic environments, simulations, virtual laboratories, and project-based activities~\cite{konak2014using,pirker2020virtual}. 
These approaches are particularly relevant because effective software development requires not only programming knowledge but also debugging, testing, reviewing, collaboration, and process-oriented decision-making~\cite{highsmith2013adaptive}. In professional practice, these responsibilities are distributed across distinct roles: developers, reviewers, testers, managers, and maintainers~\cite{jimenez2009challenges}. Role-based learning can make this division of labor visible to novices, helping them understand complex software-development workflows as a coordinated system rather than a single undifferentiated task~\cite{stuanciulescu2016technology}. Prior work suggests that assigning novices realistic roles can improve engagement and support the practical application of theoretical knowledge~\cite{hidalgo2023challenges}. Relatedly, project- and role-based approaches expose novices to real-world practices such as requirements negotiation, teamwork, and quality assurance~\cite{ceh2023application}. These findings suggest that novices may better understand complex workflows by participating from a situated role rather than merely observing outcomes. Building on this insight, our work extends role-based, immersive learning to agentic AI-supported software development, enabling novices to learn agentic workflows through active participation in a realistic AI-supported development process rather than passive observation.
\begin{figure*}[t]
    \centering
    \includegraphics[width=0.92\textwidth]{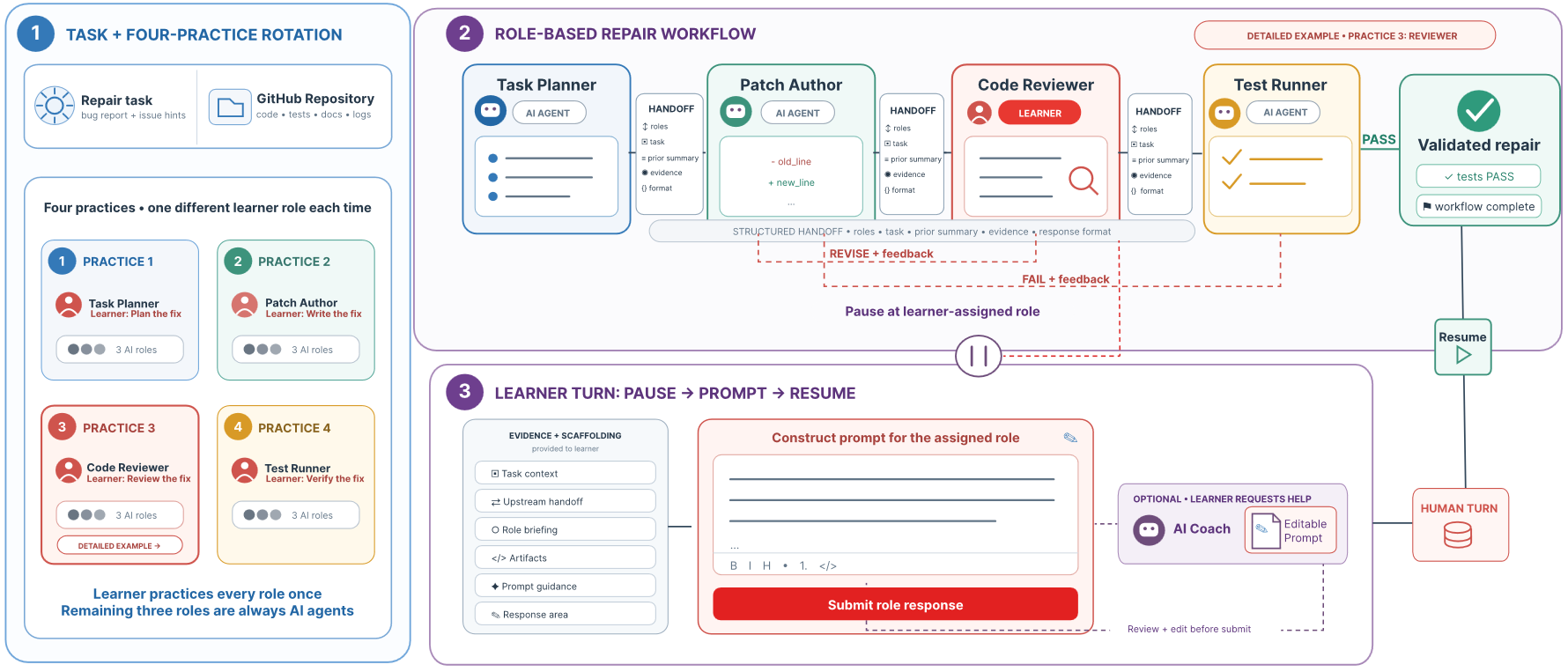}
    \caption{Overview of \textit{AgentForge}. Learners rotate through four software-engineering roles while AI agents perform the others; the detailed example shows the workflow pauses for the learner’s role-specific response (e.g., code reviewer) before continuing to revision, testing, or completion.}
    \label{fig:workflow}
\end{figure*}

\section{Methodology}
\label{sec:methodology}
We designed and evaluated \textit{AgentForge}, a role-immersion system that enables novices to practice software repair with AI agents. In a within-subject study, participants completed four role-based practice sessions, assuming one development role while AI agents performed the others. We aim to understand:
\begin{itemize}
    \item How do novices interact with AgentForge while completing software repair tasks?
   \item What task performance and learning outcomes do novices achieve?
   \item What are novices' perceptions of and experiences with AgentForge?
\end{itemize}

\subsection{System Design}
\label{sec:system-design}
\textit{\textbf{(1) Multi-Agent Learning Practices.}} Inspired by prior multi-agent frameworks~\cite{qian2024chatdev, hong2024metagpt}, \textit{AgentForge} decomposes the code-repair process into a four-role workflow: \textit{Task Planner}, \textit{Patch Author}, \textit{Code Reviewer}, and \textit{Test Runner}, as shown in Figure~\ref{fig:workflow}.

The \textit{Task Planner} translated the issue description into a structured repair plan. Participants in this role summarized the proposed repair approach, identified the likely root cause, named relevant files or components to inspect, and defined acceptance checks. This practice emphasized problem decomposition before implementation, as vague or unspecified plans can lead downstream agents to make broad or unreliable code changes.

The \textit{Patch Author} converted the repair plan into a concrete code change. Participants described the intended implementation, identified modified files, proposed a patch or specific change, and specified completion criteria. This practice helped novices connect debugging hypotheses to concrete edits while keeping AI-assisted changes minimal, traceable, and aligned with the original issue.

The \textit{Code Reviewer} evaluated the proposed patch for correctness, regression risk, missing edge cases, and clarity. Participants decided whether the patch should be approved for testing or returned for revision, and provided supporting review notes. In the workflow, an \texttt{APPROVE} decision moved the repair to testing, whereas a \texttt{REVISE} decision routed it back to the Patch Author with feedback. This practice emphasized critical evaluation of AI-generated code rather than uncritical acceptance.

The \textit{Test Runner} assessed whether the submitted patch was supported by validation evidence. A repair was considered validated only if the patch applied successfully and the targeted tests passed. 
Participants interpreted evidence (e.g.,  patch-application and test-execution results), then decided whether the workflow had enough evidence to finish or should return for revision. This practice helped novices distinguish plausible code from validated code, an important distinction in AI-assisted software development.\\
\textit{\textbf{(2) Role-Based Workflow Scaffolding and Metacognitive Support.}}
As explained above, we decomposed the code-repair process into a four-role workflow. 
To support novices within this workflow, \textit{AgentForge} scaffolded their participation through structured role handoffs and role-specific guidance. Each handoff identified the upstream and downstream actors and roles, the current task, a summary of the preceding output, the available evidence, and the response format expected from the next role. 
By making role responsibilities, information flow, and coordination explicit, these handoffs helped novices interpret upstream artifacts, produce structured outputs for downstream agents, and follow the progression of the software-repair workflow.
Each role was also accompanied by a step briefing that described its purpose, relevant software-engineering concepts, recommended procedures, success criteria, and a structured prompt template. 

When the workflow reached the novice's assigned role, execution paused, and the interface presented the issue description, relevant code files, the recent workflow transcript, the upstream handoff described above, the role briefing, available artifacts, role-specific evidence, and a response area.
To complete the response, novices constructed a prompt directing the AI system to carry out their assigned role's action. The template provided optional hints and structural guidance while allowing novices to determine which instructions, evidence, and constraints to include. This prompt-construction activity was intended to externalize novices' planning and monitoring of the role's responsibilities. After submission, the novice's response was recorded as a human turn, and the workflow resumed (see Figure~\ref{fig:workflow}).

\begin{figure*}[t]
    \centering
\includegraphics[width=\textwidth]{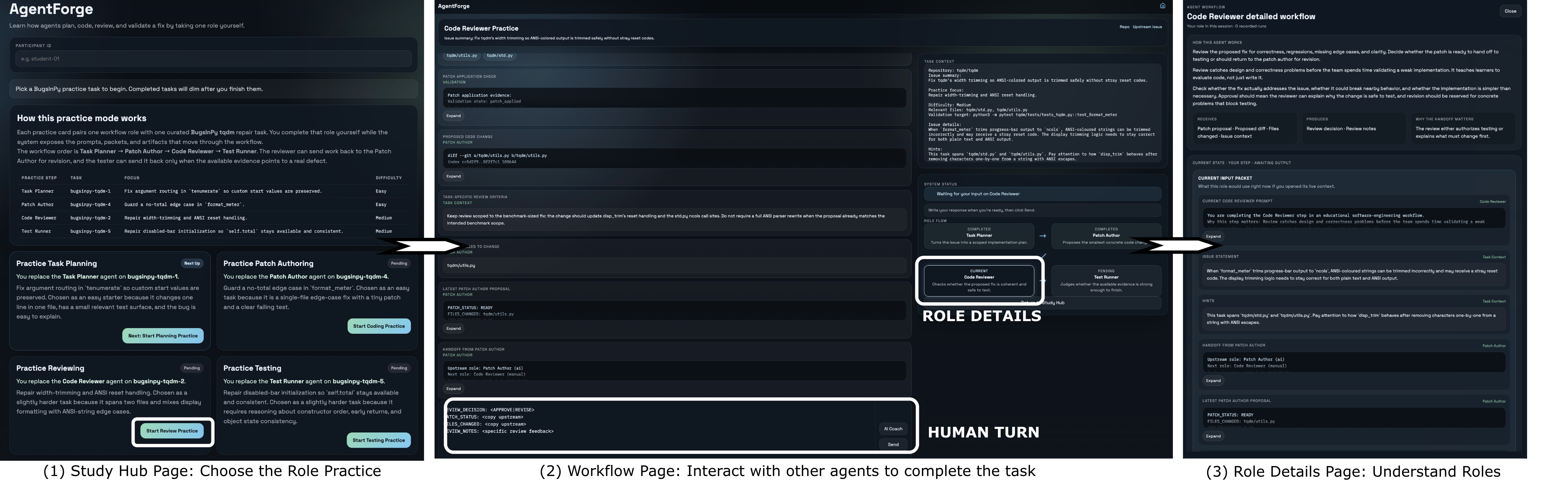}
    \caption{Three-level participant interface: (1) the study hub presents four role-based practice tasks; (2) the workflow page shows the agent workflow, task artifacts, files, handoffs, response area, and status, while enabling interaction with the AI Coach and other agents; and (3) the role-details page presents the assigned role’s purpose, prompt, inputs, outputs, context, hints, and relevant files. Arrows indicate the navigation path.}
    \label{fig:interface}
\end{figure*}

Novices could also request assistance from the AI Coach, which generated an editable prompt. This feature supported novice-initiated help seeking, a recognized form of metacognitive regulation, without removing novices' agency to accept, revise, or ignore the suggested response.\\
\textit{\textbf{(3) Interface and Implementation.}}
The participant interface comprised three navigational levels, as shown in Figure~\ref{fig:interface}. Participants began at the study hub, which presented four practice tasks corresponding to the Task Planner, Patch Author, Code Reviewer, and Test Runner roles. Selecting a task opened the main workflow page, where participants could inspect the agent workflow, role instructions, inter-role handoffs, repository files, response field, and current workflow status. Clicking the assigned role opened a third window displaying a detailed view of the role's purpose, inputs, outputs, prompt, issue description, hints, handoffs, and relevant files. Relevant-file links opened the corresponding files directly in the IDE.

The backend was implemented in Python with FastAPI, and AI roles were executed through the GPT-5 API. The system recorded session state and events, including role assignments, messages, handoffs, support requests, workflow transitions, patch applications, test runs, and completion status. 

\subsection{Task Design}
\label{sec:task-design}
\textit{\textbf{(1) Code Repair Tasks.}}
\label{sec:task-design-repair}
The study used four code-repair tasks from the BugsInPy benchmark version of the \texttt{tqdm/tqdm} repository, which has been adopted in prior software-engineering research and education~\cite{fang2023four,cassee2025barriers}. 
The tasks were selected to represent realistic maintenance activities while remaining accessible to novices with limited software-development experience. They covered argument routing, missing-value handling, ANSI-aware display trimming, and object-state initialization.
The Task Planner and Patch Author tasks were classified as easy because they involved localized, single-file repairs. The Code Reviewer and Test Runner tasks were classified as medium because they required reasoning across multiple files or about display-width edge cases, constructor ordering, and state consistency.

We used BugsInPy as both the source of repair tasks and the basis for patch evaluation. For each task, the system loaded the benchmark issue context, base repository version, relevant file metadata, and suggested test command, and presented participants with the bug report, optional hints for learners, the buggy repository version, the files related to the bug, the test command used to check the repair, and criteria describing what a successful fix should satisfy. For validation, the system applied the current patch artifact to a clean checkout of the corresponding BugsInPy benchmark repository and ran the task-specific benchmark test command.
The role--task mapping was fixed for all participants to keep the practice sequence consistent. \\
\textit{\textbf{(2) Human Surveys.}}
We administered the pre- and post-surveys through Qualtrics\footnote{\url{https://www.qualtrics.com/}}.
Participants completed a pre-survey before using \textit{AgentForge} and a post-survey after completing all practice tasks. The pre-survey collected background information, including software development experience, prior use of AI coding tools, and familiarity with multi-agent collaboration. It also included six Likert-style items assessing participants' understanding of software development activities and nine Likert-style items assessing their understanding of planning, coding, review, testing, instructing AI agents, verifying outputs, and participating confidently in an authentic code-repair workflow.

The post-survey repeated the nine workflow-understanding items to measure pre--post changes. It also included 19 multiple-choice knowledge questions covering software repair workflows, role responsibilities, agent outputs, and the interpretation of code-validation results. In addition, the post-survey contained eight Likert-style items assessing system helpfulness, twelve items assessing usability and user experience, one item assessing the system's usefulness for software development training, and six open-ended questions about role clarity, learning outcomes, confusing elements, and suggested improvements.

\subsection{Experiment}
\label{sec:study-procedure}
This study has been approved by our local Institutional Review Board. 
We recruited 37 participants from a senior-level Software Engineering course, all of whom had foundational software engineering knowledge but limited experience with real-world code repair or multi-agent systems. Participants then completed a post-survey and received 0.5 bonus points toward their final course grade as compensation. 
The majority (97.3\%) of the participants self-reported having less experience than advanced programmers. 
Participants also reported prior use of AI programming tools, most commonly ChatGPT (86.5\%), Claude (83.8\%), and Gemini (64.9\%).

The study was conducted during a regular class session and consisted of a pre-survey, practice sessions, and a post-survey. After a brief introduction and system demonstration, participants completed the pre-survey.
Participants then completed up to four practice sessions corresponding to the \emph{Task Planner}, \emph{Patch Author}, \emph{Code Reviewer}, and \emph{Test Runner} roles, in any order and at their own pace. 
In each session, we instructed the participants to review the issue description, role-specific instructions, upstream handoffs, and repository artifacts before submitting a structured response. 
They could optionally request assistance from the AI Coach. 
After submission, they observed subsequent agent interactions and routing decisions. 
Each session ended when the workflow completed successfully, defined as the final patch applying correctly and all targeted tests passing, or when the maximum number of iterations was reached. Sessions that did not complete successfully after six reroutes were terminated and recorded as incomplete.

In total, we collected interaction logs from 37 participants. Of these, 31 completed all four role practices, while six completed two or three. The final dataset comprised 183 practice-session logs, including retries following incomplete or failed attempts. For each participant–role combination, we retained only the last attempt for analysis, yielding 140 unique participant–role observations.
\subsection{Data Analysis} \label{sec:data-analysis}
\textit{\textbf{(1) Interaction Analysis.}}
For each practice, we characterized human--AI collaboration using the numbers of human and AI turns, workflow reroutes, AI Coach requests, active human response time, and completion duration. 
Human turns were participant-submitted responses, AI turns were agent-generated responses, and reroutes were workflow decisions that returned the process to an earlier role. Active human response time was calculated as the sum of the intervals between system requests and the corresponding participant submissions, whereas completion duration was measured from practice initiation to workflow completion. Differences across practices were examined using Friedman tests, followed by Holm-adjusted paired Wilcoxon signed-rank tests.

We also linked each AI Coach request to its corresponding human turn and classified participants into three Coach-use strategies: \textit{No Coach Use}, \textit{Coach Use Without Revision}, and \textit{Coach Use With Revision}, based on whether they used the Coach and whether they modified its generated drafts. Participant responses were descriptively scored from 0 to 100 using an exploratory deterministic rubric assessing format compliance, task alignment, role-specific evidence, and actionable reasoning, with each dimension worth up to 25 points~\cite{cohn2025cotal,wang2026autoscore}. Scores were generated through a Claude Sonnet 5-assisted process and independently verified by one researcher, with 89.1\% agreement between the automated and human assessments.
In addition, we examined associations between Coach-use strategy and response quality, workflow-understanding gains, post-test scores, task completion, and patch similarity using Kruskal--Wallis tests with Holm-adjusted post hoc comparisons.\\
\textit{\textbf{(2) Performance and Learning Analysis.}}
Performance was evaluated using task-completion rate and patch similarity. A task was considered complete when the generated patch applied successfully and passed the targeted benchmark tests. Differences in completion rates across practices were examined using Cochran's $Q$ test.
Patch similarity was measured using signed token-edit F1. The generated patch and the benchmark-provided ground-truth patch were represented as added and deleted code-token edits, with edit direction preserved. A score of 1 indicated identical edits, while missing or mismatched edits reduced the score. Missing or unparsable patches were excluded. Differences across practices were tested using Friedman tests, followed by Holm-adjusted paired Wilcoxon signed-rank tests.

Learning outcomes were assessed using the post-knowledge test and nine matched pre--post workflow-understanding items. Correct post-test responses were converted to percentage scores, and pre--post differences in workflow understanding were examined using paired Wilcoxon signed-rank tests.
\\
\textit{\textbf{(3) Perceived Usefulness.}}
Likert-scale survey responses were coded from 1 to 5, and construct scores were calculated by averaging the corresponding items. Cronbach's $\alpha$ was used to assess the internal consistency of multi-item constructs. Spearman correlations examined associations between perceived Coach helpfulness, ease of interaction, confidence in system outputs, and learning support and participants’ observed behavior and performance. Holm correction was applied within related families of statistical tests. Open-ended responses were analyzed by one researcher using descriptive inductive coding.
\section{Results}
\label{sec:results}

\subsection{How do novices interact with AgentForge while completing software repair tasks?}
\label{sec:results-rq1}

We found the Task Planner and Patch Author practices each achieved a 100\% completion rate, followed by the Test Runner at 93.5\%. The Code Reviewer had the lowest observed completion rate at 80.6\%. However, among the 31 participants who attempted all four practices, completion rates did not differ significantly across roles.

\begin{table}[t]
    \centering
    \caption{Completion and AI Coach use by task-role practice.}
    \label{tab:rq1-completion}
    \scriptsize
    \begin{tabular}{lrrrr}
    \toprule
    Condition & Attempted & Completed & AI Coach Used (\%) \\
        \hline
        Task Planner & 37 & 37 & 62.2\% \\
        Patch Author & 36 & 36 & 61.1\% \\
        Code Reviewer & 36 & 29 & 61.1\% \\
        Test Runner & 31 & 29 & 41.9\% \\
        \bottomrule
    \end{tabular}
\end{table}


\begin{table}[t]
    \centering
    \caption{Participant-level rounded average interaction measures by task-role practice.}
    \label{tab:rq1-interactions}
    \scriptsize
    \setlength{\tabcolsep}{3pt}
    \renewcommand{\arraystretch}{0.7}
    \begin{tabular}{lcccc>{\centering\arraybackslash}p{1cm}}
        \toprule
        Measure & \makecell{Task\\Planner} & \makecell{Patch\\Author} & \makecell{Code\\Reviewer} & \makecell{Test\\Runner} & $p_{\mathrm{adj}}$ \\
        \midrule
        Human turns & 1 & 1 & 2 & 1 & {\scriptsize$<$}.001 \\
        AI turns & 3 & 3 & 6 & 3 & {\scriptsize$<$}.001 \\
        AI Coach requests & 1 & 1 & 2 & 0 & .021 \\
        Reroutes & 0 & 0 & 2 & 0 & {\scriptsize$<$}.001 \\
        Human response time (min) & 3 & 4 & 5 & 2 & .030 \\
        Completion time (min) & 4 & 4 & 7 & 3 & .004 \\
        \bottomrule
    \end{tabular}
\end{table}

Interaction patterns varied notably across the four task-role practices. For the Task Planner, Patch Author, and Test Runner roles, novices generally completed the task in a single pass, requiring only one human intervention. By contrast, the Code Reviewer role produced the heaviest interaction workload, with the highest average number of human turns (2), AI turns (6), and reroutes (2), as well as the longest overall completion time (7 min) as shown in Table~\ref{tab:rq1-interactions}. This reflected novices revisiting earlier workflow steps when final tests failed, rather than spending substantially more time on each turn, suggesting that Code Reviewer practice imposed greater coordination demands through more frequent revise-and-resubmit cycles. 
In addition, the Patch Author and Code Reviewer roles required longer human response time, indicating that these practices remained cognitively demanding even when overall interaction burden differed.

Furthermore, the AI Coach was used frequently throughout the study, with usage not varying significantly across the four practices; Test Runner showed the numerically lowest usage rate (41.9\%). This pattern may reflect the procedural nature of the practice: the testing workflow itself provided direct validation feedback, so novices may have had fewer occasions to request explanatory or drafting support from the Coach.
Beyond these practice-level patterns, we also examined how novices used the Coach across all human turns. Of the 214 human turns collected, 84 (39.3\%) did not involve an AI Coach, 111 (51.9\%) used an AI Coach without revision, and 19 (8.9\%) used an AI Coach with revisions.
As introduced in the data analysis, we further evaluated response quality using rubric scores across these three groups. The Coach Use Without Revision group received the highest mean rubric score (94.6), compared with 58.0 for the No Coach Use group and 79.1 for the Coach Use With Revision group.
At the participant level, nine participants never used the AI Coach, 16 used it without revision, and 12 used it with revision. After Holm correction, Coach-use strategy was not significantly associated with response quality, changes in participants' self-reported understanding of the agent workflow (pre- to post-survey), post-test knowledge scores, task-completion rate, patch similarity to reference solutions, or the rate at which generated patches passed the targeted tests.

\subsection{What task performance and learning outcomes do novices achieve?}
\label{sec:results-rq2}
We then evaluated task performance using test pass rate and patch similarity between the generated and reference patches. Because a practice was considered complete only when the patch applied successfully and passed the targeted benchmark tests, completion rate was equivalent to test pass rate. As shown in Table~\ref{tab:rq1-completion}, the Test Runner only achieved a completion rate of 93.5\%. 
As discussed in the previous Section, participants also used the AI Coach less frequently during the Test Runner practice. Together, these patterns may suggest miscalibrated confidence: participants may have believed they could determine whether the code was valid without additional support, even when their validation judgments were incorrect.
Moreover, among participants who completed all four practices, patch similarity differed significantly across roles ($p_{\mathrm{adj}} < .001$). Mean signed token-edit F1 scores were .68 for Task Planner, .52 for Code Reviewer, .39 for Patch Author, and .65 for Test Runner, with Patch Author showing the lowest similarity. Most pairwise differences remained significant after Holm correction, except between Task Planner and Test Runner. This pattern may indicate that the greater implementation autonomy afforded in the Patch Author role led to more variable solutions than the more structured Task Planner and Test Runner roles.

We then evaluated learning outcomes using self-reported pre- and post-survey measures and an objective knowledge test. After using the system, participants reported a stronger understanding of structured software repair workflows, with the mean score increasing from 3.8 to 4.3 ($p_{\mathrm{adj}} < .001$, $r = .851$). They also reported greater understanding of the value of multiple AI agents, increasing from 3.3 to 4.3 ($p_{\mathrm{adj}} < .001$, $r = .946$), and greater preparedness to participate in software repair workflows, increasing from 3.4 to 4.0 ($p_{\mathrm{adj}} = .004$, $r = .908$). Scores for understanding the sequence of roles and handoffs, instructing AI agents, and verifying AI-generated outputs also increased descriptively, although these changes were not statistically significant.
Performance on the objective knowledge test was generally high, with participants answering an average of 82.6\% of questions correctly. 
They performed well on questions about the purposes of planning, review, testing, and handoffs in a multi-agent repair workflow. However, they were less accurate when applying specific decision rules in determining whether to approve a patch with only stylistic concerns or one that fixes the bug using an inefficient implementation. 
Interpreting test evidence was the weakest area, with only 20.0\% of participants answering correctly.

Together, these findings suggest that the system strengthened participants' overall understanding of software repair workflows and agent collaboration. However, future support should move beyond broad workflow explanations and provide greater scaffolding for nuanced review and validation decisions.

\subsection{What are novices’ perceptions of and experiences with AgentForge?}
\label{sec:results-rq3}
Participants rated \textit{AgentForge} positively for supporting their understanding of AI-assisted software repair, including planning, review, testing, role coordination, collaboration with AI agents, and the overall repair process (mean = 4.1, $\alpha = .86$). Specifically, at the item level, most participants agreed that the system helped them understand why code review matters (91.4\%), supported learning of the software-development process (91.0\%), and helped them understand collaboration with AI (88.6\%); 82.9\% considered the system useful for software-development training, and 80.0\% rated the AI Coach as helpful.
Ratings were also positive for usability-related aspects, including workflow clarity, handoff clarity, instruction clarity, novice manageability, explanation quality, and ease of interaction (mean = 3.8, $\alpha=.82$). 
The lower usability composite score was driven primarily by two concerns: only 60.0\% of participants considered the system manageable for novices with limited development experience, and only 45.7\% found the interface user-friendly. Nevertheless, 62.9\% agreed that the system explained concepts clearly. 
Participants further reported positively perceived outcomes, including willingness to use the system again, perceived task effectiveness, and learning support (mean = 3.9, $\alpha=.74$).
To assess consistency between self-reported perceptions and observed performance, we conducted exploratory correlations between the two; none remained significant after Holm correction, including Coach helpfulness with observed Coach requests, ease of interaction with human response time, confidence in system output with patch similarity, and perceived learning support with workflow-understanding gain.

In terms of perceived benefits and challenges, Task Planner was most often identified as the easiest role (45.7\%), while Code Reviewer was most often identified as the most challenging (25.7\%). 
The most common learning gain (65.7\%) was an understanding of how the four roles work together in an end-to-end code repair workflow. The most common source of confusion, by contrast, was unclear instructions or response expectations (51.4\%), followed by interface density (28.6\%).
Participants most frequently requested more detailed examples, tutorials, or feedback, along with a more readable layout. 
These results suggest that participants valued the immersive learning structure but struggled to translate it into the precise responses expected.
This may be because of the 130 human turns submitted following a Coach request, 85.4\% exactly reproduced the Coach draft, and 88.5\% were at least 95\% text-identical to it. 
This indicates that many nominally human submissions reflected acceptance of AI-generated content rather than critical engagement or deeper understanding. 
Overall, participants found the system helpful for understanding how to collaborate with AI in software repair workflows.
\section{Discussion and Conclusion}
\label{sec:discussion-conclusion}
With agentic AI support, participants achieved high task-completion rates and reported stronger understanding of software repair and AI-agent collaboration. However, code review and patch validation remained the most challenging aspects. Novice education should therefore place greater emphasis on these skills: although code generation is becoming increasingly accessible, ensuring correctness, quality, and maintainability still requires human judgment.
These findings also inform the design of future AI education tools. Rather than simply restricting AI use, educational systems should make intermediate artifacts, agent handoffs, and decision points visible and actively involve learners in guiding and evaluating AI decisions. Such support should promote active reasoning by asking learners to justify decisions, critique AI suggestions, identify missing evidence, and revise proposed responses. Because many participants accepted AI-generated content with little or no revision, scaffolding should gradually fade as learners gain experience to reduce cognitive offloading. Future systems could also present intentionally incomplete outputs so that learners can practice detecting errors and deciding whether to accept, revise, or regenerate results. Clearer examples, concise instructions, and less dense interfaces may further reduce unnecessary cognitive load.
Overall, \textit{AgentForge} shows that role-based immersive learning can help novices view software development as a coordinated process rather than an isolated coding task. The key challenge is not only helping learners complete AI-assisted work, but also developing their ability to evaluate outputs and remain accountable for final decisions.

\bibliography{aaai2027}


\end{document}